\input{aipcheck}

\documentclass[
    ,final            
  ]
  {aipproc}

\layoutstyle{8x11double}

\begin{document}

\title{The MINERvA Neutrino Interaction Experiment}


\author{Richard Gran}{
  address={University of Minnesota Duluth}
}



\begin{abstract}
 The MINERvA neutrino interaction experiment in the NuMI beam
at Fermilab will measure several aspects of neutrino interactions
in the few GeV energy region.
We will make cross section and form factor measurements using 
a fine-grained fully active scintillator (CH) target, and also
investigate nuclear effects on neutrino interactions as well as 
hadron rescattering using integral nuclear targets made of helium,
carbon, iron, and lead.  We will improve or add to existing measurements
and address some specific questions that are important for current 
and upcoming neutrino oscillation experiments.  This paper reports on the progress toward
the design, construction, and calibration of the detector, which
we expect will get its first neutrino beam in 2009.
\end{abstract}

\maketitle


\section{Overview of the MINERvA experiment}

MINERvA has a broad physics measurement program.
We will measure neutrino interaction cross sections both 
inclusive and exclusive, and with very high statistics,
using the NuMI neutrino beam at Fermilab.  In addition,
we will simultaneously study nuclear effects on neutrino interactions
using several integral nuclear targets.

The two most important pieces of the MINERvA experiment are
the fully active, fine-grained inner tracker region and the
upstream nuclear target region.  The tracker layers are made
of triangular shaped bars of extruded scintillator with 3.3 cm
base and 1.7 cm height, the latter sets the plane thickness.  
By analyzing the energy sharing between
these interlocking triangles, we will have 3 mm 
position resolution for reconstructed tracks.  The scintillator
planes are arranged in UXVX sequences to give stereo 3D reconstruction 
of multi-track events.  Each plane is 
a 2.2 meter wide hexagon; 120 planes make this inner tracker region
and another 76 planes are described in the next two paragraphs.

The same tracking layers continue upstream
of the inner tracker to the nuclear target region where layers of
carbon, iron, and lead are interspersed.  Because of the nearly identical
tracking layers, we will have a direct comparison of interaction rates and 
hadron production and rescattering between these nuclear targets.  
Though the tracking environment
will not be identical, we will also analyze data from a liquid Helium target region
just upstream from the solid nuclear targets.

Other features of the detector include a downstream electromagnetic
calorimeter (ECAL) region with 2 mm thick lead sheets
and a hadron calorimeter (HCAL) with one inch steel plates between scintillator planes.
Behind this is the existing MINOS near detector which we will use as a 
muon spectrometer.  There are also electromagnetic and hadron calorimeter regions 
along the sides of the detector to measure particles leaving the neutrino
vertex or scattered at large angles, making the detector nearly hermetic.  
Taken together, the MINERvA detector has excellent
vertex, tracking, and calorimetric coverage for most products produced
by neutrino interactions.

The fully active region will have approximately three tons fiducial volume
of plastic scintillator (CH) in the inner tracker and another 6.2 tons of
material in the nuclear target region, when the plastic, carbon, iron, and 
lead are combined.  For the expected one year running
in the current NuMI low energy beam (4 x 10$^{20}$ protons on target) and
three years running in the medium energy beam (12 x 10$^{20}$ POT), we will
accumulate over nine million charged-current events in the fully active fiducial volume.  
Within this we expect to have a sample
of 800,000 quasielastic and 1.6 million resonance events, with another 
2 million in the resonance-DIS transition region and over 4 million in the deep 
inelastic scattering region.  Total event samples from the nuclear targets
will range from 400,000 to 2.5 million, giving very high statistics for 
those comparisons. 

The collaboration successfully completed a full module (two scintillator planes)
prototype and full DAQ system tests during Spring and Summer 2007.
The MINERvA experiment also completed its Department of Energy
Critical Decision Three (CD3) review, and approval was granted in November 2007.

\section{Physics Capabilities}

Because the MINERvA measurement program is so broad, I will only highlight
a few of the physics capabilities.  More complete details on all the 
measurements that have been considered quantitatively can be found in chapter
two of the MINERvA Technical Design Report, available on the web
\cite{TDR}.  Topics not covered here include coherent pion 
production \cite{Gallagher2007},
deep inelastic scattering, structure functions, and quark-hadron duality.

\subsection{Flux}

Several absolute cross section measurements depend on an accurate model of 
the neutrino flux through the MINERvA detector.  This flux model will be based on
a combination of measurements.  Following the MINOS experience, we will have 
neutrino data from multiple beam configurations \cite{NUMI}, which allow some 
ability to disentangle the flux model from the cross section model.
There are results coming out from the first run of Fermilab E-907 (MIPP) \cite{MIPP}
with the hadron spectrum off of the MINOS target.  
There is also a proposed upgrade program for MIPP \cite{MIPPupgrade} which will provide additional
improvements to their hadron production measurements.   Finally, there is an initiative
to improve information from the muon monitors to more directly gain information
from rate of pion decays to muons and neutrinos \cite{KoppNuInt07}.
When this information is incorporated into the analysis, we expect to know
the flux to the 5\% level for most of the 1 to 20 GeV region, with a little less
constraint at the edges of the beam's focusing peak.

\subsection{Test beam calibrations}

A beam test of the MINERvA detector design and components is being prepared.  
We will expose a 40 plane mini-MINERvA detector where each plane is 107 x 107 cm square,
which is 63 strips wide instead of the full 128 strips.  These planes will
have the same UXVX sequence and tracking capabilities as the full MINERvA detector.

To test the tracker, electromagnetic, and hadron calorimetry regions, the lead
and steel absorber will be removable.  This will allow us to test all the basic
configurations and combinations.  In addition to all ECAL and all HCAL configurations,
two examples of combinations we are likely to test are:
10 tracker + 20 ECAL + 10 HCAL layers or 20 ECAL + 20 HCAL. 

The top priority for the test beam effort is to expose the detector to hadrons
in as similar a momentum range as possible and obtain a calibration of the
calorimetric quantities such as total visible energy deposit and its fluctuations. 
For quasielastic and resonance 
interactions, the pion momentum distribution is dominated by protons and pions 
at and below 500 MeV/c.  At around this momentum, it is already likely that the
hadron will undergo an inelastic interaction rather than range out.  On the other
hand, the spectrum from the low hadron invariant mass interactions has a tail
going up to a few GeV while high invariant mass ``deep inelastic scattering'' events
have a much higher momentum spectrum.  

The beamline that will host this measurement is Fermilab's Meson Test Beam Facility.
This beam has recently been upgraded to provide a usable pion rate down to
momenta of 1 GeV/c.  The design of a tertiary beam and beam selection components
is underway; we expect it will give pions to 300 MeV/c.  The beam will also deliver
protons, electrons, and muons for other tests and calibrations.  We are scheduled to
run in Fall 2008.

\subsection{Quasielastic interactions}

Quasielastic interactions represent one of the most important subsamples 
in the MINERvA data.  Because of the simple two-body kinematics and the MINERvA
tracker design, we expect to have full kinematic reconstruction down to very low $Q^2$.
The quasielastic cross section is the anchor for all cross sections around 1 GeV,
where the single pion production cross section starts to turn on, but well below where 
the  inclusive cross section is dominated by multi-hadron production and well measured 
above 50 GeV.  Except possibly for the tau appearance signature at CNGS, all other 
recent, current, and upcoming neutrino oscillation experiments will depend on an
accurate model or constraint of this cross section for their measurement.

Because of the expected knowledge of the flux, MINERvA will be able to provide a
measurement of this exclusive cross section based on the observed rate.  In addition,
the tracking abilities and high statistics will allow for a study of the axial vector
form factor, including both the extraction of the $M_A$ parameter as well as investigation
into deviations from the simple dipole form.  In MINERvA, such studies will be limited
by the eventual systematic errors, including the knowledge of the flux spectrum, 
the muon range calibration, and the hadron energy calibrations mentioned above.  
We will also study the more abundant but more complicated single pion final states,
including resonance production.

At this time, there appears to be a few puzzles about the quasielastic interaction,
which MINERvA will be in a good position to address.  The extracted value of $M_A$ 
from K2K \cite{K2K} and MiniBooNE \cite{MiniBooNE} appears to be slightly higher than
expected based on previous deuterium bubble chamber measurements.  At this time,
it is not certain whether one or more nuclear effects for oxygen
or carbon are mismodeled, or if there is some other fundamental change in the underlying form factor.
It may also simply be some systematics inherent in how these measurement techniques
differ from those used with the exquisite kinematic information available from
the bubble chamber data.

There is also some concern about whether the parameter $M_A$ is adequate to 
reproduce both the absolute cross section as well as the shape of the $Q^2$ distribution
simultaneously, especially across the range of nuclei and energies of interest today.

A second puzzle is the very low $Q^2 < 0.2$ (GeV/c)$^2$ region.  Separate from the 
bulk of the $Q^2$ distribution, the models currently in use do not do a good job
reproducing the data.  Recently, MiniBooNE  has presented a new 
model parameter that allows them to reproduce this feature of the $Q^2$ spectrum
for their oscillation analysis\cite{MiniBooNE}.

The MINERvA measurements will benefit from the combination of enormous statistics, 
improved constraint on the flux, excellent tracking down to 
very low momentum, and good energy reconstruction.  MINERvA covers
a range of energies that overlap with the K2K and MiniBooNE measurements on the
low end, which include or are just above the region where NOvA and T2K will expect
their oscillation signatures.  This energy range continues up to tens of GeV
where some kinds of nuclear effects should become negligible.  Finally, the 
integral nuclear targets will further allow us to isolate and identify potential
effects due to the nuclear environment.


\subsection{Nuclear reinteractions}

Another topic MINERvA will address is nuclear reinteractions.  
Hadrons produced in the final state of any of these interactions with a
single nucleus must escape the nuclear environment before they will be seen
in the detector.  Protons will rescatter, occasionally producing another
pion or kicking out a neutron.  More importantly, pions can undergo a variety
of interactions, from simple scattering to charge exchange to absorption.

Some experiments, such as MINOS, are dependent on hadron calorimetry in 
order to reconstruct and observe distortions in the neutrino energy spectrum.
Therefore, a mis-modeling of the visible hadron energy is an important systematic
effect.  Other experiments, such as those which use the water Cerenkov technique,
depend on the kinematic reconstruction of the event and their ability to 
select and model quasielastic events.  
In this case, the identification of the recoil proton from other protons or pions
coming from non quasielastic events in oxygen nuclei is important.  
Finally, the upcoming searches for electron appearance will have to contend with 
a significant background from neutral pions, which are produced directly 
and also indirectly by these reinteractions.

MINERvA will combine the excellent tracking and high statistics of the inner
tracker made of scintillator (CH) with the event rate and track multiplicity 
observed in the tracking layers downstream from the integral nuclear targets.
Downstream from the inner tracker is an electromagnetic calorimeter which will
force the conversion of and provide good information about the gamma rays from
neutral pion decays.

\section{Outlook}

During Spring 2008 and into the summer we will construct a multi-plane
tracking prototype which will include inner tracker as well as 
electromagnetic calorimeter sections.  

The full detector installation will happen in 2009, followed by operation
in the low energy NuMI beam along with the end of the MINOS run.
We expect our initial results from this run.   A changeover
to the medium energy beam required for the NOvA experiment is currently expected
to happen in 2011-2012, followed by another three years of operation.








\begin{theacknowledgments}
  The MINERvA experiment is supported by the U.S. Department of Energy, the                   
U.S. National Science Foundation, the Office of Special                         
Accounts for Research Grants of the University of Athens,                       
Greece, DAI-PUCP and CONCYTEC from Peru, the CNPq and CAPES                     
from Brazil, CONCYTEG and CONACYT from Mexico and Apoyo a la                   
Investigacion Cientifica from the Universidad de Guanajuato.
\end{theacknowledgments}





\end{document}